\documentclass{ptapap}

\author{Emese Plachy}[CSFK]
\author{L\'aszl\'o Moln\'ar}[CSFK]
\author{Attila B\'odi}[CSFK,SZTE]
\author{Marek Skarka}[CSFK,CZ]
\author{\'Aron L. Juh\'asz}[CSFK]
\author{\'Adam S\'odor}[CSFK]
\author{P\'eter Klagyivik}[CSFK,IAC,UL]
\author{R\'obert Szab\'o}[CSFK]
\affil[CSFK]{Konkoly Observatory, MTA CSFK, Budapest, Hungary}
\affil[SZTE]{Department of Experimental Physics and Astronomical Observatory, University of Szeged, Hungary}
\affil[CZ]{Astronomical Institute, Czech Academy of Sciences, Fri\v{c}ova 298, 251 65 Ond\v{r}ejov, Czech Republic}
\affil[IAC]{Instituto de Astrof\'isica de Canarias, 38205 La Laguna, Tenerife, Spain}
\affil[UL]{Universidad de La Laguna, Dept. de Astrof\'isica, 38206 La Laguna, Tenerife, Spain}

\title{K2 Photometry of RR~Lyrae Stars}

\begin{document}

\maketitle

\begin{abstract}
Thousands of RR~Lyrae stars have been observed by the \textit{Kepler} space telescope so far. We developed a photometric pipeline tailored to the light variations of these stars, called the Extended Aperture Photometry (EAP). We present the comparison of our photometric solutions for Campaigns 0 through 6 with the other pipelines available, e.g., SAP/PDCSAP, K2P2, EVEREST, and others. We focus on the problems caused by instrumental effects and the detectability of the low-amplitude additional modes. 
\end{abstract}

\section{Introduction}

Space-based photometry during the last few years led to new discoveries in the pulsation of RR~Lyrae stars at the micro-magnitude level. Some of the low-amplitude additional modes have already been explained: subharmonic frequencies appear due to the nonlinear phenomenon of period doubling \citep{szabo2010, buchler}, while the frequencies of $f_{\rm x}$ mode near the $0.61$ period ratio belong to non-radial oscillations \citep{dziem}, and overtone modes can also be present with very low amplitudes \citep{benko}. However many of these low-amplitude frequencies are still puzzling \citep{molnar2012}. High-data quality is essential to study these additional modes, which can be achieved with space-based photometry or very extended ground-based photometry, like OGLE \citep{netzel,smolec,prudil}. 

The \textit{K2} mission differs from the \textit{CoRoT} and the original \textit{Kepler} missions in the sense that it observes multiple large fields providing a significantly larger sample of high-quality, quasi-continuous RR~Lyrae light curves that forms the basis for statistical analysis. These analyses have the potential to recover connections of the different phenomena in RR~Lyrae stars that may lead to the explanation of all the low-amplitude modes or even the Blazhko effect. 

More than four thousand RR~Lyrae stars have been observed in the \textit{K2} mission so far. However, the two-wheel observing mode presents unique challenges before the data can be analysed. The attitude changes of the telescope introduce systematic variations in the photometry of the stars that can distort or mask the intrinsic light variations of the stars. Various pipelines have been developed to eliminate the effects of these spacecraft motions that work well for various signals (transits, low-amplitude, and/or long period variations), but not for RR~Lyrae stars. The automated methods cannot handle the timing of attitude correction manoeuvres because it is in the range of the pulsation periods and the sharp features that they cause in the light curves resemble the typical RRab light curve shapes, especially around maximum light. 

In this paper we demonstrate on an example star that scientific results are sensitive to the \textit{K2} photometric pipelines.

\section{Methods} 

The existing pipelines provide solutions for two tasks: the photometry itself and corrections of the instrumental effects. We investigated the RR~Lyrae light curves derived with the following pipelines: Simple Aperture Photometry (SAP) and Pre--search Data Conditioned SAP \citep[PDCSAP,][]{vancleve}, the Self--flatfielding correction method \citep[K2SFF,][]{vanderburg}, \textit{K2} Pixel Photometry \citep[K2P$^2$,][]{lund}, the \textit{K2} Variability Catalog \citep[K2VARCAT,][]{armstrong}, EPIC Variability Extraction and Removal for Exoplanet Science Targets \citep[EVEREST,][]{luger}, and \textit{K2} Systematics Correction \citep[K2SC,][]{aigrain}. The latter method only provides corrections for the SAP and PDCSAP light curves. We found that these methods may fail for RR~Lyrae stars for two reasons: the apertures are too tight and/or the corrections can not distinguish between instrumental and RR~Lyrae variations. In many cases, the raw photometry gives better results without the corrections. However, we demonstrate that K2SC is a promising correction method in the case of RR~Lyrae stars, if it is applied to a suitable photometry. 

\begin{figure}
\begin{center}
\includegraphics[width=1.\textwidth]{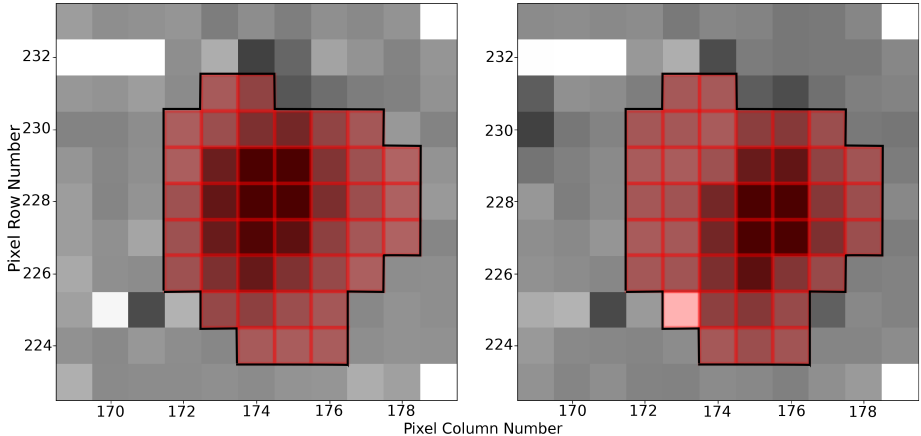}
\caption{Extended apertures of EPIC 211048310 in the two extrema of photocenter positions.}
\label{fig:mask}
\end{center}
\end{figure}

The need for large apertures for large amplitude pulsators was already recognized in the original \textit{Kepler} mission. In the \textit{K2} mission it is even more essential due to the continuous change of the photocenter positions that can reach 2--3 pixels at the edges of the field of view. Therefore, we developed the Extended Aperture Photometry (EAP) method \citep{plachy} where the basic idea is to extend the aperture until it fully contains the PSF of the target in the two extrema of the photocenter positions (Fig.~\ref{fig:mask}). The method handles the targets individually and iteratively finds the best aperture for each; therefore, it is time-consuming. On the other hand, with the extended apertures, the systematics are significantly reduced. The correction frequencies decrease below the noise level when we apply the K2SC method to the light curves. Since the apertures are large, possible contaminations can cause trends or additional variation in the average flux that we eliminate with a spline-smoothing algorithm. This method provided us better quality light curves than the other pipelines in the majority of cases of our examined sample (of \textit{K2} Guest Observer proposal GO4069 for Campaign 4).

We created EAP light curves for Campaigns 0 to 6 for more than five hundred stars. A detailed description and data publication are in preparation (Moln\'ar et al. in prep.). Here we present only one example to show how useful the EAP photometry can be in the search of low amplitude modes.

\section{Example}

\begin{figure}
\begin{center}
\includegraphics[width=\textwidth]{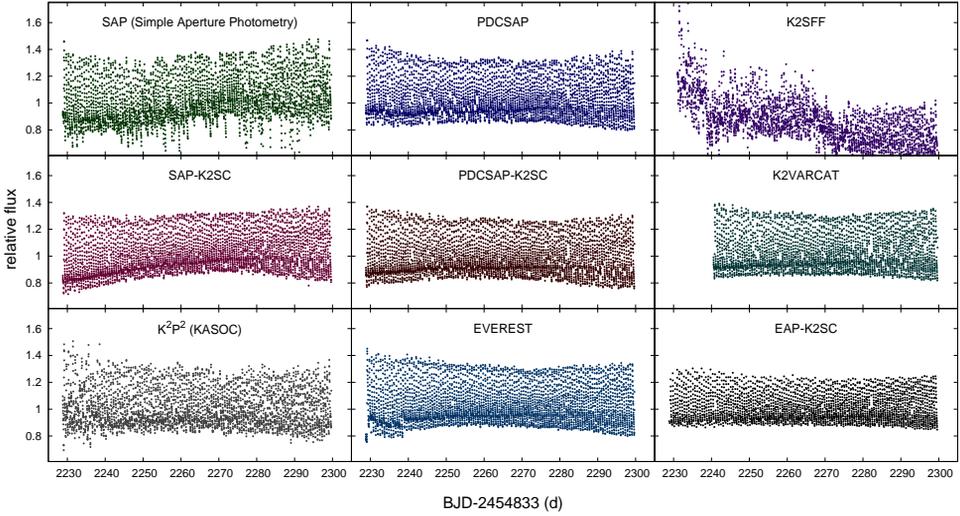}
\caption{Light curve solutions of EPIC 211048310 from the different pipelines.}
\label{fig:lc}
\end{center}
\end{figure}

\begin{figure}
\begin{center}
\includegraphics[width=\textwidth]{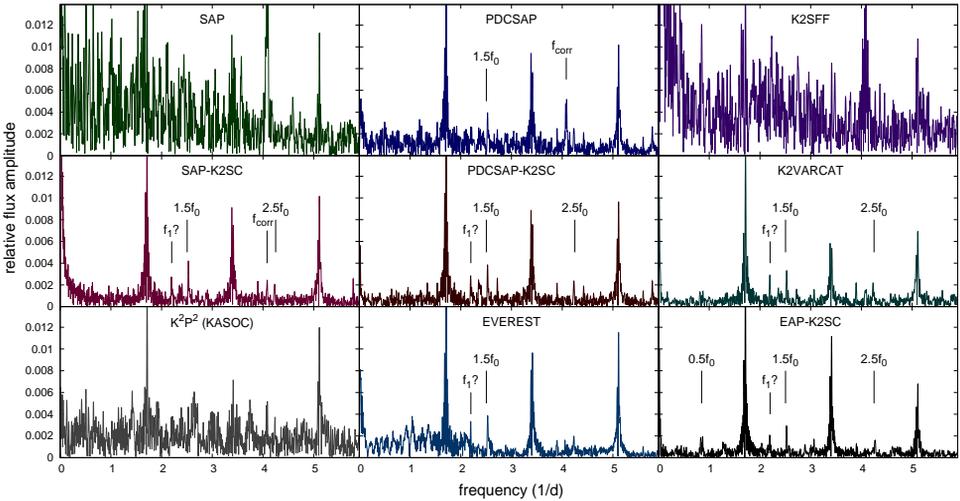}
\caption{The low-amplitude region of the residual spectra of the light curves of EPIC 211048310 from the different pipelines after prewhitening with $f_0=1.6998$\thinspace c/d and its harmonics. The subharmonics (0.5, 1.5 and 2.5$f_0$), the attitude correction frequency, $f_{\rm corr}\approx 4.06$\thinspace c/d, and an additional frequency at $f_1= 2.1955$\thinspace c/d are marked.}
\label{fig:res}
\end{center}
\end{figure}

We chose a fundamental-mode RRab star from Campaign 4 to demonstrate the differences of the photometric pipelines. EPIC 211048310 ($K_p=15.530$\thinspace mag) is a Blazhko star with a long modulation period that is not covered in \textit{K2} data. Period doubling (half-integer frequencies) is expected in the Blazhko stars. We compared not only the light curves but the residual spectra as well as to look for the signs of period doubling and other possible additional modes. In Fig.~\ref{fig:lc}, we can see that most of the light curves suffer from various issues (trends, jumps, noise, missing parts). If we investigate the low-amplitude frequencies in Fig.~\ref{fig:res}, we may notice that while the subharmonic frequencies are present in several spectra at $\sim1.5f_0$ and $\sim2.5f_0$, the $0.5f_0$ subharmonic appears only in the EAP photometry with a significant amplitude. 

We note that this is only one example, and differences in the light curve quality vary from star to star. For example, the EVEREST method is able to produce high-quality light curves in about half of the cases, while in the other half, it removes the RR~Lyrae variation during correction. Hopefully further versions of EVEREST will solve this problem.

\section{Conclusions}

The RR~Lyrae data from the \textit{K2} mission require special treatment. Enormous efforts are needed to achieve the best photometric solution. We believe that with the EAP method, we can improve light curves significantly compared to other pipelines published at the time of writing this paper; however, our method is not easy to automate. The first EAP data release will be available soon.

\acknowledgements{
This work has used \textit{K2} targets selected and proposed by the RR~Lyrae and Cepheid Working Group of the Kepler Asteroseismic Science Consortium (proposal number GO4069). Funding for the \textit{Kepler} and \textit{K2} missions is provided by the NASA Science Mission directorate. This project has been supported by the Lend\"ulet LP2014-17 Program of the Hungarian Academy of Sciences, and by the NKFIH K-115709, PD-121203 and PD-116175 grants of the Hungarian National Research, Development and Innovation Office. EP, LM and \'AS were supported by the J\'anos Bolyai Research Scholarship of the Hungarian Academy of Sciences. Additionally, MS acknowledges the financial support of the Czech Grant GA \v{C}R 17-01752J and the support of the postdoctoral fellowship programme of the Hungarian Academy of Sciences at the Konkoly Observatory as host institution. The research leading to these results have been supported by the \'UNKP-17-3 program of the Ministry of Human Capacities of Hungary.}

\bibliographystyle{ptapap}
\bibliography{eplachy_K2}

\end{document}